\documentclass[sigconf]{nimeart}
\AtBeginDocument{%
  }

\setcopyright{cc}
\copyrightyear{2025}
\acmYear{2025}
\acmDOI{}
\acmConference[NIME '25]{International Conference on New Interfaces for Musical Expression}{June 24--27,2025}{Canberra, Australia}
\acmISBN{}

\begin{document}

\title{Two Sonification Methods for the MindCube}

\author{Fangzheng Liu}
\authornote{Both authors contributed equally to this research.}
\email{fzliu@media.mit.edu}
\orcid{0000-0003-0202-0892}
\affiliation{%
  \institution{MIT Media Lab}
  \city{Cambridge}
  \state{MA}
  \country{USA}
}

\author{Lancelot Blanchard}
\authornotemark[1]
\email{lancelot@media.mit.edu.edu}
\orcid{0000-0003-1580-3116}
\affiliation{%
  \institution{MIT Media Lab}
  \city{Cambridge}
  \state{MA}
  \country{USA}
}

\author{Don D. Haddad}
\email{ddh@mit.edu}
\orcid{0000-0001-6420-3219}
\affiliation{%
  \institution{MIT Media Lab}
  \city{Cambridge}
  \state{MA}
  \country{USA}
}

\author{Joseph A. Paradiso}
\email{joep@media.mit.edu}
\affiliation{%
  \institution{MIT Media Lab}
  \city{Cambridge}
  \state{MA}
  \country{USA}
}

\renewcommand{\shortauthors}{Liu et al.}

\begin{abstract}
  In this work, we explore the musical interface potential of the MindCube, an interactive device designed to study emotions. Embedding diverse sensors and input devices, this interface resembles a fidget cube toy commonly used to help users relieve their stress and anxiety. As such, it is a particularly well-suited controller for musical systems that aim to help with emotion regulation. In this regard, we present two different mappings for the MindCube, with and without AI. With our generative AI mapping, we propose a way to infuse meaning within a latent space and techniques to navigate through it with an external controller. We discuss our results and propose directions for future work.
\end{abstract}



\maketitle

\section{Introduction and Previous Work}
Miniature and handheld music controllers can provide intuitive and expressive means for musical interaction \cite{torre2016hands, bonifacic2020noise}, often leveraging innovative designs.

The "Kibo" \cite{amico2020kibo} is a MIDI controller featuring a simplified tangible user interface, designed entirely from wood. It comprises eight geometric extractable solids that users can manipulate to trigger note events and control various musical parameters. The device is an intuitive and tactile learning tool, aiming for enhanced music education. The "Accordiatron" \cite{gurevich2020accordiatron} is another novel MIDI controller inspired by the traditional concertina. It translates the performer's gestures into MIDI data, allowing for flexible mapping to various musical parameters. The combination of discrete and continuous sensory outputs provides the subtle expressiveness necessary for interactive music performance.
The "AirSticks" \cite{trolland2022airsticks} is another gestural musical instrument that integrates Inertial Measurement Units (IMUs) to enable performers to trigger and manipulate sound events in real-time through expressive gestures. This wireless device captures both discrete actions, such as striking motions, and continuous movements, allowing for nuanced control over various musical parameters.
The "AirSticks" successfully showed the power of commercial IMU for capturing striking and fluid motions in real time.
With modern IMUs getting smaller, lower-noise, and lower power consumption, we can design more compact and efficient music controllers without sacrificing the fidelity of the captured motion data.
The "CD-synth" \cite{chwalek2019cd} is a compact, wireless digital synthesizer designed for expressive musical performance. Its disc-shaped form allows performers to freely rotate and reorient the instrument, utilizing non-contact light sensing to modulate sound parameters. Equipped with sensors that detect rotation, orientation, touch, and proximity, the CD-Synth manipulates audio filters and effects applied to preset wavetables.

Some research uses off-the-shelf interactive systems to create musical interaction. Wong, Yuen \& Choy \cite{wong2008designing}, for example, use the Nintendo Wii Controller to develop an interactive music performance system. By employing analytical techniques to study motion data captured by the controller, the system maps detected gestures to musical expression. This approach leverages a low-cost and readily available game controller to create an engaging musical interface. Most of these interfaces, due to the large amount of data they produce through their sensors, are also interesting candidates to control and manipulate Machine Learning (ML) models and generative systems. ML has been extensively used to design music interfaces, mostly through the learning of explicit mappings between controls and sound characteristics \cite{Fiebrink2009AMF, 10.1145/2502081.2502214, Fried2013, 10.1145/2557500.2557544}. To enable further sonic exploration, \textit{latent space exploration} has been proposed as a way to create an implicit or explicit mapping between the internal representation of an ML model and a set of chosen sound characteristics. Previous work \cite{campos_generative_2018, chen_flat_2022, scurto_soundwalking_2023, nime2023_32} has been mostly focused on the unconditional exploration of latent spaces for sound generation. To enable better control, some work has focused on guiding the latent space exploration with a given set of conditioning signals. Bitton et al. \cite{bitton_assisted_2019}, for example, enable the sampling of a 3-dimensional latent space learnt by an Adversarial Auto-Encoder (AAE) to generate new musical samples that comply with a set of given characteristics (e.g., timbre or playing technique). Bretan et al. \cite{bretan_deep_2017} use nearest neighbor search to automatically continue the musical input of a live performer. Only limited work has focused on real-time latent space exploration with strict conditioning for audio generation as we do here.


In this paper, we propose a way for the MindCube, a device we designed in previous work \cite{liu2024mindcube}, to be used as an interactive music controller. Resembling a fidget cube toy commonly used for stress and anxiety relief, the MindCube offers a more compact form factor compared with the above-mentioned work--only $3.3 cm \times 3.3 cm \times 3.3 cm$, which is significantly smaller than many existing controllers. Its design allows it to be comfortably held and operated with one hand, enhancing its portability and user-friendliness. Despite its small size, the MindCube is equipped with various interactive inputs, including tactile buttons, a rolling disk, a joystick, and a 9 DoF IMU, which can detect the controller's attitude in hand, providing a rich set of controls for musical expression.

\section{Instrument Design}
\subsection{Hardware}
\begin{figure}[ht!]
  \centering
  \includegraphics[width=\linewidth]{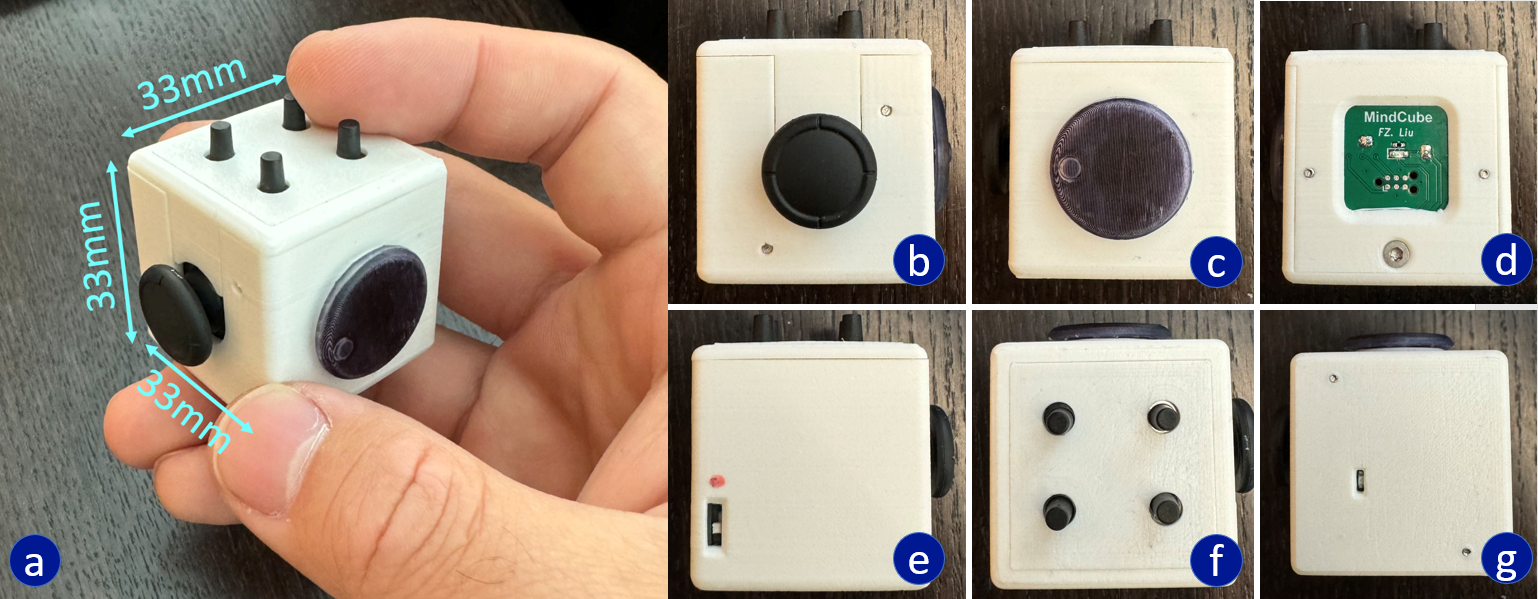}
  \caption{(a) A MindCube in hand and configurations of each side:
  (b) a joystick, (c) a rolling disk, (d) the charging indicator and programming port, (e) the power switch and linear vibration motor (on the inside), (f) tactile switches, (g) an LED indicator}
  \label{mindcubeonahand}
\end{figure}

The MindCube design resembles a fidget cube toy commonly used for stress and anxiety relief \cite{aspiranti2022using, biel2017fidget, da2018identifying, driesen2023tools}.
A MindCube is a miniature (3.3cm $\times$ 3.3cm $\times$ 3.3cm) cubic interactive device that is easy to hold with one hand, which makes it ideal for playful interaction. Each side of the MindCube has various interactive inputs, including four tactile buttons, a small rolling disk, and a joystick, as shown in Figure \ref{mindcubeonahand}.

\begin{figure}[ht!]
  \centering
  \includegraphics[width=\linewidth]{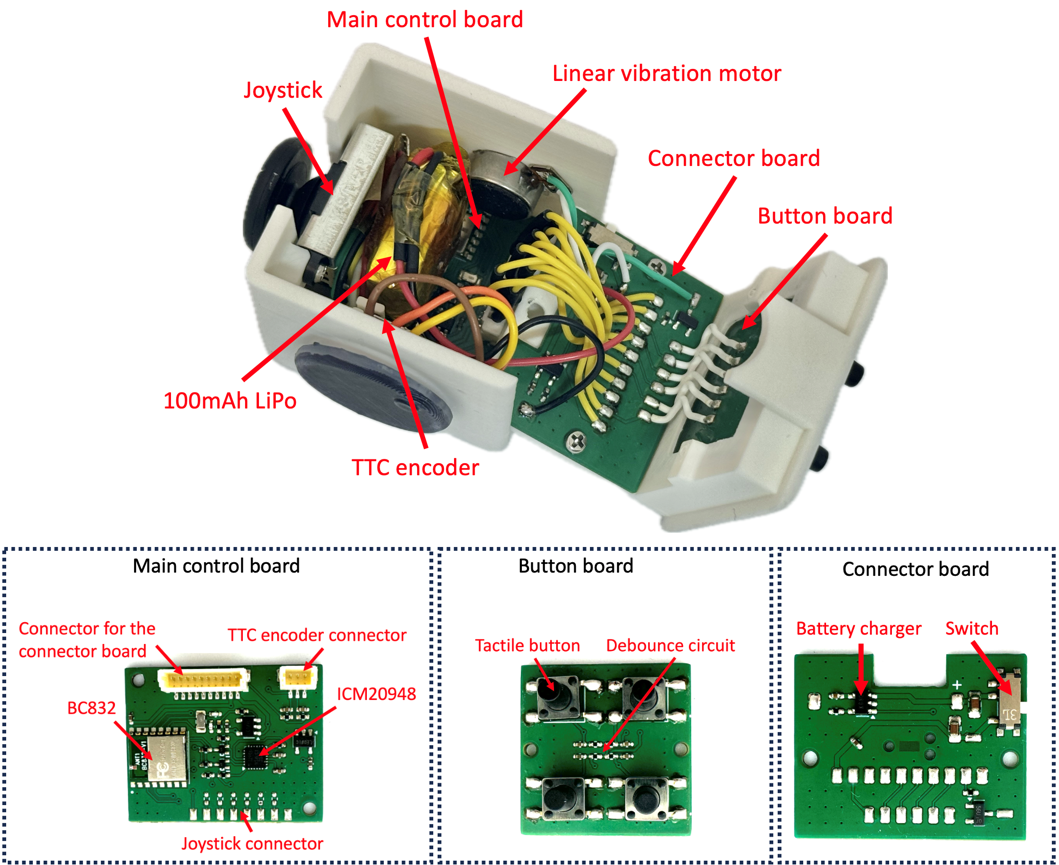}
  \caption{Three PCBs inside a MindCube.}
  \label{cubeinside}
\end{figure}

The rolling disk is connected to a mouse scroll wheel encoder, and the SoC measures the pulses from it to detect rolling distance and directions. The inside of a MindCube is shown in the Figure. \ref{cubeinside}.

The MindCube contains three PCB boards, each dedicated to a specific function. The main control board manages all control and communication processes. It is equipped with an nRF52832 (ARM Cortex-M4, Nordic) Bluetooth Low Energy (BLE) system-on-chip (SoC). Additionally, the board includes an ICM-20498 9-DoF IMU, which captures 3-axis accelerometer, gyroscope, and magnetometer data. This data enables real-time tracking of the MindCube’s orientation while in the user’s hand. Mounted on the opposite side of the main control board, the button board features four tactile buttons with debounce circuits to eliminate mechanical switch bouncing. The connector board serves as a bridge between the button board and the main control board. It also integrates a programming port for flashing firmware onto the SoC, a charging port for the Li-Po battery, and a slide switch to turn the MindCube on or off. To prevent accidental power-off, the switch handle is lower than the surface of the MindCube body.
The system diagram in Figure \ref{systemdiagram} details the MindCube system structure.

\begin{figure}[ht!]
  \centering
  \includegraphics[width=\linewidth]{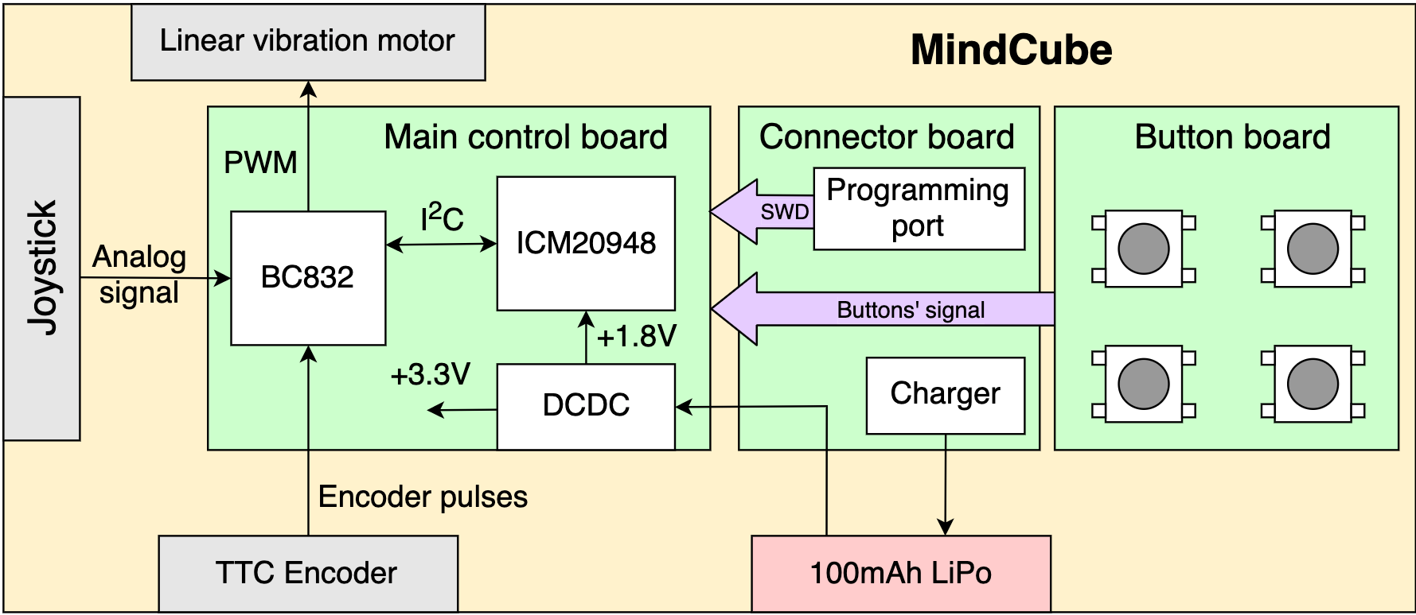}
  \caption{The system diagram of the MindCube.}
  \label{systemdiagram}
\end{figure}

The MindCube is powered by a 100 mAh Li-Po battery, providing up to more than three hours of battery life during continuous data transmission. Additionally, a linear vibration motor is mounted inside, which can be programmed to deliver various haptic feedback patterns. The motor is controlled via pulse-width modulation (PWM).

\subsection{Firmware and communication}
The MindCube firmware is developed using the Arduino framework. The SoC continuously reads sensor measurements, packages the data into MindCube packets, and transmits them via Bluetooth Low Energy (BLE) at a rate of 20 Hz. To ensure reliable and unambiguous packet framing, each packet is COBS (Consistent Overhead Byte Stuffing) encoded.
A Python-based front-end application running on a MacBook receives the data over BLE, decodes the packets, and processes the information for various applications. One potential use case is analyzing the data to study users' real-time emotional states \cite{woodward2020ifidgetcube}. In this paper, we explore the data sonification applications.

In the following sections, we describe two musical mapping approaches utilizing the MindCube, with and without generative AI. The AI-driven approach explores the potential of using the MindCube's data to estimate the user's current emotional state and generates music as a proxy for emotion regulation. Although we do not formally prove here that data from the MindCube can detect a user's emotional status, our working hypothesis is that increased interaction with the MindCube may indicate heightened stress levels, while decreased interaction could suggest a more relaxed state. In response, our AI model generates calming music when high activity is detected and stimulating music to engage the user when low activity is detected. We aim to prove the MindCube's potential to accurately detect users' emotions in future work. To contrast with our AI approach, we also offer a non-AI musical mapping, where the MindCube facilitates expressive musical performances through a handcrafted modular synthesizer mapping.

\section{AI-generated Music Mapping}
In this AI-powered musical mapping, we collect the sensor data and generate loud, high-energy music when the user activity is low, and quiet, low-energy music when the user activity is high. This experimental mapping aims to engage the user continuously and hopes to be able to regulate the user's emotional state over time. An implementation of this mapping can be found on GitHub\footnote{\url{https://github.com/mitmedialab/mindcube-rave}}.

\subsection{Model Architecture}

To create an AI-based musical mapping for the MindCube, we make use of the RAVE (Realtime Audio Variational autoEncoder) model architecture \cite{caillon2021ravevariationalautoencoderfast}. RAVE is based on a Variational Auto Encoder (VAE) trained on accurately reconstructing audio files by encoding them to a latent distribution, before decoding them into audio files. The appeal of using this architecture within musical instruments stems from its capabilities to perform this autoencoding faster than real time. Following the VAE mathematical notation, we consider music as a continuous signal $x$ sampled from an underlying data distribution $p_\text{data}(x)$. The RAVE model allows us to learn a latent representation of dimension 4 ($z \in \mathbb{R}^4$) that captures meaningful musical features while allowing for efficient reconstruction. The encoder and decoder of our VAE are neural networks that are modeled by $q_\phi(z|x)$ and $p_\theta(x|z)$.

The diverse sensors and input devices of the MindCube offer us the opportunity to explore the latent space of the model in a fun and interactive way. In order to explore this latent space in real time, we use Latent Diffusion \cite{rombach_high-resolution_2022} to generate latent codes that are then passed through the RAVE decoder to reconstruct audio\footnote{RAVE Latent Diffusion is implemented at \url{https://github.com/moiseshorta/RAVE-Latent-Diffusion}}. Specifically, we model the latent space traversal as a stochastic process, where a latent variable $z_T \sim \mathcal{N}(0, I)$ undergoes a sequence of denoising steps following a learned reverse diffusion process. Our denoising process is denoted as $z_{t-1} = z_t + \epsilon_\psi(z_t, t)$, where $\epsilon_\psi$ is a neural network trained to predict and remove noise at each step $t$. The final latent code $z_0$ is then decoded using $p_\theta(x|z_0)$ to synthesize the corresponding audio. This approach enables smooth and structured navigation of the latent space, allowing the MindCube to generate expressive musical transformations in real time.

\subsection{Model Mapping}

We then have to generate latent codes that align with the inputs of the musical instrument, in order to create an enjoyable and coherent mapping. To do so, we make use of \textit{Classifier-Free Guidance} (CFG) \cite{ho_classifier-free_2021}, a technique that allows us to modulate the generation process by conditioning on specific features. In our case, we train the model with CFG using the \textit{Root-Mean-Square} (RMS) value as a conditioning signal, which serves as a proxy for the perceived loudness and energy of the generated audio. We use RMS as our main metric to generate contrasting high-energy and low-energy music, since loud, high-energy music generally exhibits a high RMS while quieter, low-energy music typically has a low RMS. We condition the Latent Diffusion Model by projecting our RMS value to a 1024-dimensional embedding which is fed to multiple cross attention layers.

During training, the model uses conditional dropout to learn both an unconditional distribution and a joint distribution over latent variables and their corresponding RMS values. At inference time, we can then use the following conditional score function:
$$
\nabla_z \log p_\theta(z|c) = (1-\gamma) \nabla_z \log p_\theta(z) + \gamma \nabla_z \log p_\theta(z|c)
$$

where $c$ represents the RMS conditioning and $\gamma$ is the guidance weight that controls the strength of the conditioning. When inferring latents, we compute a real-time RMS value from the sensor input and normalize it to a range between 0 and 1. This value is then used as the conditioning variable $c$ in the CFG process, guiding the latent code generation towards outputs that match the desired spectral characteristics. The final latent code, as previously described, is passed through the RAVE decoder to synthesize audio, allowing for expressive and dynamic control over the instrument's sonic output.

\begin{figure}[h!]
  \centering
  \includegraphics[width=\linewidth]{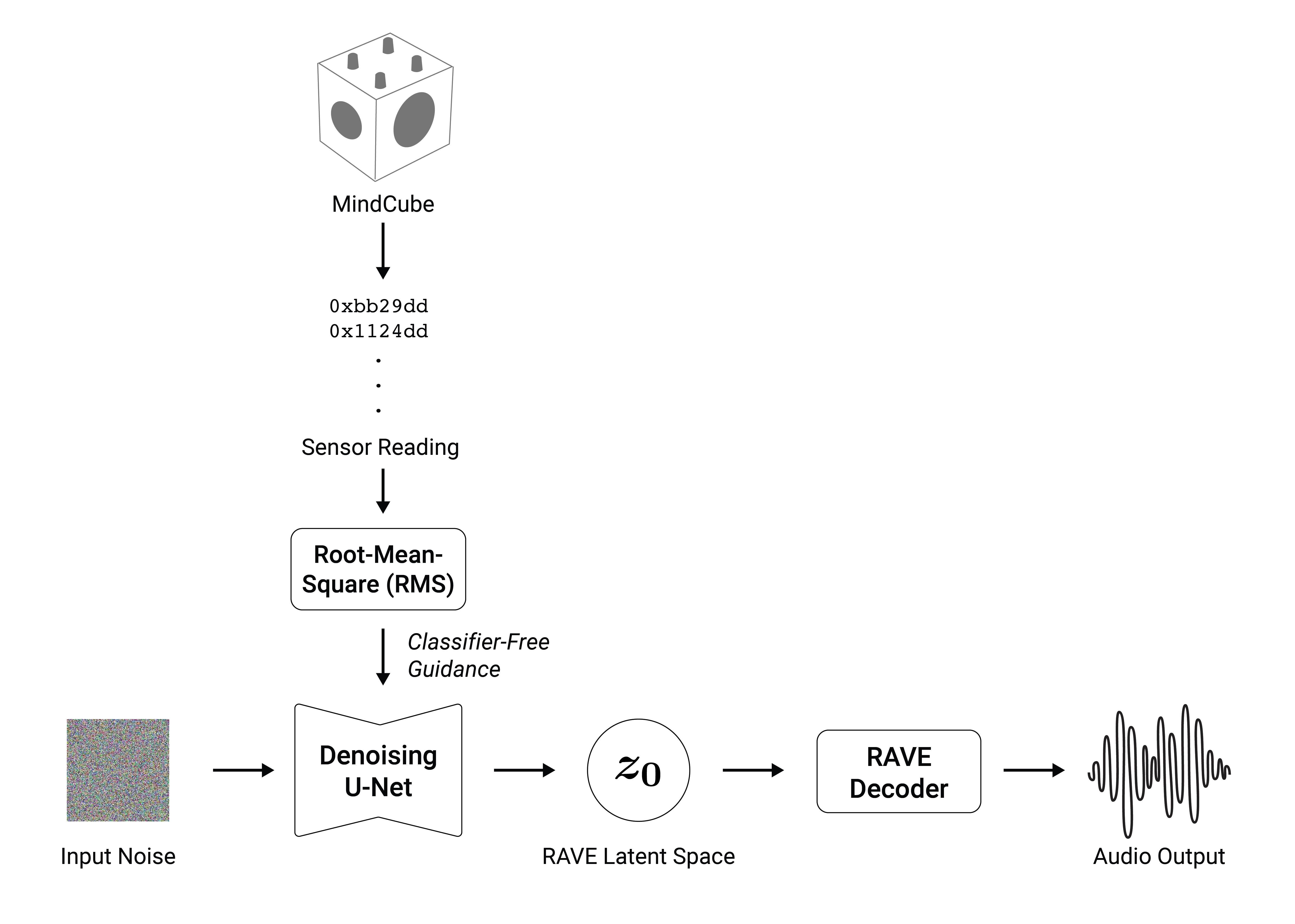}
  \caption{The architecture of the AI Music mapping for the MindCube.}
  \label{ai_arch}
\end{figure}

An overview of our architecture can be seen in Figure \ref{ai_arch}.

\vspace{-1em}
\subsection{Training}

We train our RAVE model on the Free Music Archive (FMA) dataset, in particular the \textit{``small''} subset, which contains 8,000 tracks of 30s of 8 balanced genres. On this subset of the dataset, the RMS value ranges between 0 and 0.8724. We train the model over 177 epochs until we observe the validation score going up. We use this RAVE model to encode our same dataset into latent codes with length 512 and use these latents to train our Latent Diffusion model. We then train our Latent Diffusion model with RMS as the embedding for CFG, for a total of 700 epochs.

\subsection{Real-time Generation}

We then need to embed our model into a real-time system to enable continuous music generation. This is a crucial but difficult step since both the latent diffusion and latent decoding processes are lengthy and introduce latency. To reduce this latency, we diffuse latents with a small length of 512, which, at a sampling rate of 44,100 Hz, gives us around 23 seconds of audio. Additionally, we only use 30 diffusion steps. On the M3 Max Macbook Pro that we used in our testing, the latent diffusion step took around 0.90 seconds while the latent decoding step took less than 0.15 seconds, giving us a total of around 1.05 seconds per generation. Our latency of 1.05 seconds for music generation means that we are forced to read the sensor input at a rate lower than $1/1.05 \approx 0.9524 \text{ Hz}$. Although not optimal, this latency still enables a responsive interface since it allows for the user input to be considered in under a second on average.

Our real-time generation system therefore reads the data from the input sensors every second and adds it into a buffer. Every 1.05 seconds on average, we diffuse a new latent sequence of length 512 and perform Classifier-Free Guidance to condition the diffusion on the last sensor readings. To create an effective RMS condition, we use the following formula:

\[
RMS_{cond} = \frac{1}{R} \cdot \sum_{i=1}^{16} w_i \cdot \sigma_i \ ,
\]

where $\sigma_i$ refers to the standard deviation of sensor $i$, $w_i$ is the weight for the reading of that sensor, and $R$ is a normalization factor. In other terms, this calculates the weighted standard deviation of every sensor value over a moving window, normalized to fit between the conditioning values used during training. This allows us to get a sense of the recent \textit{activity} of the MindCube, and calculate an adequate RMS value. Empirically, we observe that the accelerometer, the joystick, the buttons, and the encoder are the best indicators for manual activity. As such, we design a weight vector that favors these sensors over the others.

The latent diffusion model is also not designed to generate continuous music by default. To enable the generated music to flow naturally, we implement \textit{outpainting} to enable the latent diffusion model to generate an adequate continuation for the previous piece of music. For every generation, we use the last few latent codes played to kickstart the diffusion and diffuse only the continuation, which we decode and play. This allows for smooth transitions between every diffused latent.

\section{Music Mapping for expressive performance}
We also explore another musical mapping by using the sensor data stream from the MindCube with VCV Rack, and this is accomplished through a structured pipeline. This pipeline comprises a Python-based TCP server for real-time data parsing, sensor fusion techniques to process raw IMU data, and the mapping of computed values to control virtual voltages within the modular synthesis environment. The goal is to develop a robust system that enables real-time, motion-driven modulation of synthesis parameters.

\begin{figure}[ht!]
  \centering
  \includegraphics[width=\linewidth]{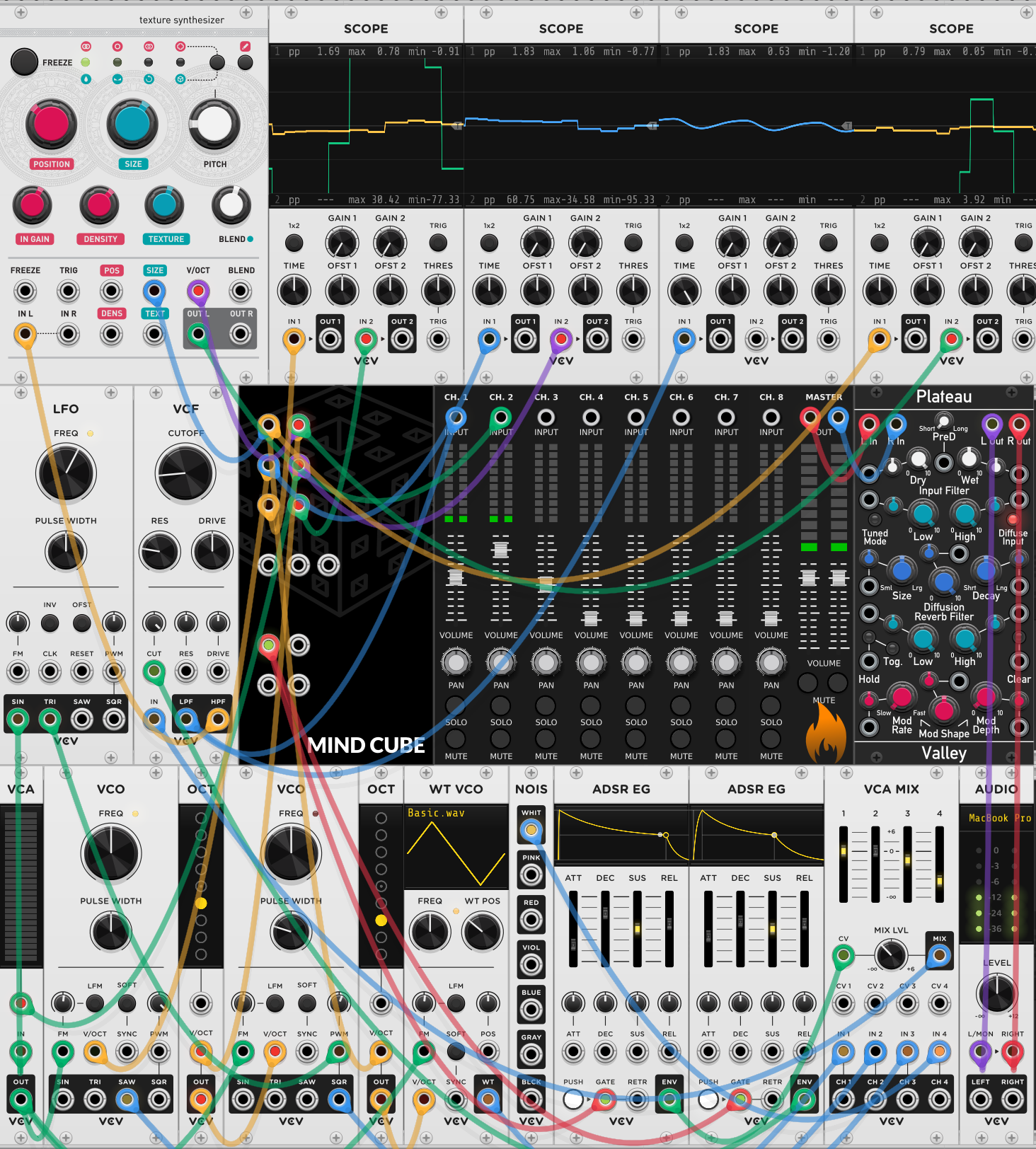}
  \caption{Live VCV Rack Patch that connects to the MindCube custom virtual module.}
  \label{systemdiagram}
\end{figure}

The MindCube streams sensor data—including accelerometer, gyroscope, magnetometer, joystick, encoder, and button states—via BLE to a Python TCP server implemented using the \texttt{bleak} library. This server listens for incoming byte data from the MindCube, parses it, and transmits the structured sensor data over a TCP socket to a custom-designed VCV Rack module, as shown in Figure \ref{systemdiagram}.
To derive meaningful control signals from the raw IMU data, a sensor fusion algorithm computes pitch and roll angles by combining accelerometer and gyroscope readings.
The TCP communication follows a local server-client model, ensuring low-latency data transmission. Data packets are formatted as comma-separated values (CSV), containing all sensor readings in a predefined order, facilitating efficient parsing and utilization within the VCV Rack environment.

The custom-designed VCV Rack module, developed in C++, interfaces seamlessly with a Python server to control the synthesizer. Operating within a dedicated thread, it continuously reads the incoming data stream, parses the received CSV strings, and converts them into floating-point values representing sensor readings. This threading approach ensures smooth integration with VCV Rack's real-time processing engine. All sensor data is normalized to fit within the modular synthesis voltage range. Various mapping strategies have been explored. For instance, computed pitch and roll values are assigned to parameters such as filter cutoff frequency and LFO rate. Joystick inputs control stereo panning and modulation index, while button states are converted into gate signals to trigger envelope generators. Additionally, encoder inputs manage step sequencing and parameter selection. This system allows users to control modular synthesis in real time by converting natural movements into dynamic sound parameters.

As AI-driven music tools continue to evolve, integrating them within established modular synthesis environments becomes increasingly relevant. Our interface, built around the BLE-enabled MindCube and mapped into VCV Rack, exemplifies how embodied interaction can extend the patching paradigm. By blending real-time sensor input with the modular workflow, we envision future systems where AI-generated modulation and user-driven control coexist fluidly--bridging algorithmic composition with the hands-on ethos of modular synthesis and patchinwg.

\section{Conclusion}
In this paper, we explore the data sonification capabilities of the MindCube, a compact, handheld interactive device designed for expressive musical performance. Its small size and various sensing modalities make it a portable and user-friendly tool. Additionally, resembling a traditional fidget cube toy, the MindCube holds a potential for real-time emotion detection. By integrating generative AI-powered real-time music generation, we aim to facilitate emotion regulation, providing different musical responses based on user interaction patterns. In the future, we will utilize the MindCube for user studies, aiming to develop an accurate model that maps its data to various emotional states. This will enhance the ground truth for our generative AI-based musical emotion regulation system. We also aim to explore sonification methodologies that can use both generative AI and modular synthesis.

\section{Ethical Standards}

There are no observed conflicts of interest. This research was conducted using discretionary funding for the hardware requirements and used lab-owned compute power for the training of the model. The Free Music Archive dataset used is distributed under the permissive \href{https://creativecommons.org/licenses/by/4.0}{CC BY 4.0} license, allowing us to use it for training and redistribution purposes.

\bibliographystyle{ACM-Reference-Format}
\bibliography{sample-references}


\begin{thebibliography}{26}


\ifx \showCODEN    \undefined \def \showCODEN     #1{\unskip}     \fi
\ifx \showDOI      \undefined \def \showDOI       #1{#1}\fi
\ifx \showISBNx    \undefined \def \showISBNx     #1{\unskip}     \fi
\ifx \showISBNxiii \undefined \def \showISBNxiii  #1{\unskip}     \fi
\ifx \showISSN     \undefined \def \showISSN      #1{\unskip}     \fi
\ifx \showLCCN     \undefined \def \showLCCN      #1{\unskip}     \fi
\ifx \shownote     \undefined \def \shownote      #1{#1}          \fi
\ifx \showarticletitle \undefined \def \showarticletitle #1{#1}   \fi
\ifx \showURL      \undefined \def \showURL       {\relax}        \fi
\providecommand\bibfield[2]{#2}
\providecommand\bibinfo[2]{#2}
\providecommand\natexlab[1]{#1}
\providecommand\showeprint[2][]{arXiv:#2}

\bibitem[Amico et~al\mbox{.}(2020)]%
        {amico2020kibo}
\bibfield{author}{\bibinfo{person}{Mattia~Davide Amico}, \bibinfo{person}{Luca~Andrea Ludovico}, {et~al\mbox{.}}} \bibinfo{year}{2020}\natexlab{}.
\newblock \showarticletitle{Kibo: A MIDI controller with a tangible user interface for music education}. In \bibinfo{booktitle}{\emph{Proceedings of the 12th International Conference on Computer Supported Education. 1: CSME}}. SCITEPRESS, \bibinfo{pages}{613--619}.
\newblock


\bibitem[Aspiranti and Hulac(2022)]%
        {aspiranti2022using}
\bibfield{author}{\bibinfo{person}{Kathleen~B Aspiranti} {and} \bibinfo{person}{David~M Hulac}.} \bibinfo{year}{2022}\natexlab{}.
\newblock \showarticletitle{Using fidget spinners to improve on-task classroom behavior for students with ADHD}.
\newblock \bibinfo{journal}{\emph{Behavior Analysis in Practice}} \bibinfo{volume}{15}, \bibinfo{number}{2} (\bibinfo{year}{2022}), \bibinfo{pages}{454--465}.
\newblock


\bibitem[Biel(2017)]%
        {biel2017fidget}
\bibfield{author}{\bibinfo{person}{Lindsey Biel}.} \bibinfo{year}{2017}\natexlab{}.
\newblock \showarticletitle{Fidget toys or focus tools}.
\newblock \bibinfo{journal}{\emph{Autism File}}  \bibinfo{volume}{74} (\bibinfo{year}{2017}), \bibinfo{pages}{12--13}.
\newblock


\bibitem[Bitton et~al\mbox{.}(2019)]%
        {bitton_assisted_2019}
\bibfield{author}{\bibinfo{person}{Adrien Bitton}, \bibinfo{person}{Philippe Esling}, \bibinfo{person}{Antoine Caillon}, {and} \bibinfo{person}{Martin Fouilleul}.} \bibinfo{year}{2019}\natexlab{}.
\newblock \showarticletitle{Assisted {Sound} {Sample} {Generation} with {Musical} {Conditioning} in {Adversarial} {Auto}-{Encoders}}. In \bibinfo{booktitle}{\emph{Proceedings of the 22nd {International} {Conference} on {Digital} {Audio} {Effects} ({DAFx}-19),}}. \bibinfo{address}{Birmingham, UK}.
\newblock


\bibitem[Bonifacic(2020)]%
        {bonifacic2020noise}
\bibfield{author}{\bibinfo{person}{Igor Bonifacic}.} \bibinfo{year}{2020}\natexlab{}.
\newblock \bibinfo{title}{Noise Machine is a tiny MIDI controller for creating music on the go}.
\newblock
\urldef\tempurl%
\url{https://www.engadget.com/noise-machine-midi-controller-231326364.html}
\showURL{%
\tempurl}


\bibitem[Bretan et~al\mbox{.}(2017)]%
        {bretan_deep_2017}
\bibfield{author}{\bibinfo{person}{Mason Bretan}, \bibinfo{person}{Sageev Oore}, \bibinfo{person}{Jesse Engel}, \bibinfo{person}{Douglas Eck}, {and} \bibinfo{person}{Larry Heck}.} \bibinfo{year}{2017}\natexlab{}.
\newblock \showarticletitle{Deep {Music}: {Towards} {Musical} {Dialogue}}. In \bibinfo{booktitle}{\emph{Proceedings of the {Thirty}-{First} {AAAI} {Conference} on {Artificial} {Intelligence}}} \emph{(\bibinfo{series}{{AAAI}'17})}. \bibinfo{publisher}{AAAI Press}, \bibinfo{pages}{5081--5082}.
\newblock
\newblock
\shownote{Place: San Francisco, California, USA}.


\bibitem[Caillon and Esling(2021)]%
        {caillon2021ravevariationalautoencoderfast}
\bibfield{author}{\bibinfo{person}{Antoine Caillon} {and} \bibinfo{person}{Philippe Esling}.} \bibinfo{year}{2021}\natexlab{}.
\newblock \bibinfo{title}{RAVE: A variational autoencoder for fast and high-quality neural audio synthesis}.
\newblock
\showeprint[arxiv]{2111.05011}~[cs.LG]
\urldef\tempurl%
\url{https://arxiv.org/abs/2111.05011}
\showURL{%
\tempurl}


\bibitem[Campos et~al\mbox{.}(2018)]%
        {campos_generative_2018}
\bibfield{author}{\bibinfo{person}{Guilherme Campos}, \bibinfo{person}{Nuno Fonseca}, \bibinfo{person}{Anibal Ferreira}, {and} \bibinfo{person}{Matthew Davies}.} \bibinfo{year}{2018}\natexlab{}.
\newblock \showarticletitle{Generative {Timbre} {Spaces}: {Regularizing} {Variational} {Auto}-{Encoders} with perceptual {Metrics}}. In \bibinfo{booktitle}{\emph{Proceedings of the 21st {International} {Conference} on {Digital} {Audio} {Effects} ({DAFx}-18),}}. \bibinfo{address}{Aveiro, Portugal}.
\newblock


\bibitem[Chen et~al\mbox{.}(2022)]%
        {chen_flat_2022}
\bibfield{author}{\bibinfo{person}{Nutan Chen}, \bibinfo{person}{Djalel Benbouzid}, \bibinfo{person}{Francesco Ferroni}, \bibinfo{person}{Mathis Nitschke}, \bibinfo{person}{Luciano Pinna}, {and} \bibinfo{person}{Patrick van~der Smagt}.} \bibinfo{year}{2022}\natexlab{}.
\newblock \bibinfo{title}{Flat {Latent} {Manifolds} for {Human}-machine {Co}-creation of {Music}}.
\newblock
\urldef\tempurl%
\url{https://arxiv.org/abs/2202.12243}
\showURL{%
\tempurl}
\newblock
\shownote{\_eprint: 2202.12243}.


\bibitem[Chwalek and Paradiso(2019)]%
        {chwalek2019cd}
\bibfield{author}{\bibinfo{person}{Patrick Chwalek} {and} \bibinfo{person}{Joe~A Paradiso}.} \bibinfo{year}{2019}\natexlab{}.
\newblock \showarticletitle{CD-Synth: a Rotating, Untethered, Digital Synthesizer.}. In \bibinfo{booktitle}{\emph{NIME}}. \bibinfo{pages}{371--374}.
\newblock


\bibitem[da~C{\^a}mara et~al\mbox{.}(2018)]%
        {da2018identifying}
\bibfield{author}{\bibinfo{person}{Suzanne~B da C{\^a}mara}, \bibinfo{person}{Rakshit Agrawal}, {and} \bibinfo{person}{Katherine Isbister}.} \bibinfo{year}{2018}\natexlab{}.
\newblock \showarticletitle{Identifying children's fidget object preferences: toward exploring the impacts of fidgeting and fidget-friendly tangibles}. In \bibinfo{booktitle}{\emph{Proceedings of the 2018 Designing Interactive Systems Conference}}. \bibinfo{pages}{301--311}.
\newblock


\bibitem[Driesen et~al\mbox{.}(2023)]%
        {driesen2023tools}
\bibfield{author}{\bibinfo{person}{Matson Driesen}, \bibinfo{person}{Joske Rijmen}, \bibinfo{person}{An-Katrien Hulsbosch}, \bibinfo{person}{Marina Danckaerts}, \bibinfo{person}{Jan~R Wiersema}, {and} \bibinfo{person}{Saskia Van~der Oord}.} \bibinfo{year}{2023}\natexlab{}.
\newblock \showarticletitle{Tools or Toys? The Effect of Fidget Spinners and Bouncy Bands on the Academic Performance in Children With Varying ADHD-Symptomatology}.
\newblock \bibinfo{journal}{\emph{Contemporary Educational Psychology}}  \bibinfo{volume}{75} (\bibinfo{year}{2023}), \bibinfo{pages}{102214}.
\newblock


\bibitem[Fiebrink et~al\mbox{.}(2009)]%
        {Fiebrink2009AMF}
\bibfield{author}{\bibinfo{person}{Rebecca Fiebrink}, \bibinfo{person}{Dan Trueman}, {and} \bibinfo{person}{Perry~R. Cook}.} \bibinfo{year}{2009}\natexlab{}.
\newblock \showarticletitle{A Meta-Instrument for Interactive, On-the-Fly Machine Learning}. In \bibinfo{booktitle}{\emph{New Interfaces for Musical Expression}}.
\newblock
\urldef\tempurl%
\url{https://api.semanticscholar.org/CorpusID:9059668}
\showURL{%
\tempurl}


\bibitem[Fran\c{c}oise(2013)]%
        {10.1145/2502081.2502214}
\bibfield{author}{\bibinfo{person}{Jules Fran\c{c}oise}.} \bibinfo{year}{2013}\natexlab{}.
\newblock \showarticletitle{Gesture--sound mapping by demonstration in interactive music systems}. In \bibinfo{booktitle}{\emph{Proceedings of the 21st ACM International Conference on Multimedia}} (Barcelona, Spain) \emph{(\bibinfo{series}{MM '13})}. \bibinfo{publisher}{Association for Computing Machinery}, \bibinfo{address}{New York, NY, USA}, \bibinfo{pages}{1051–1054}.
\newblock
\showISBNx{9781450324045}
\urldef\tempurl%
\url{https://doi.org/10.1145/2502081.2502214}
\showDOI{\tempurl}


\bibitem[Fried and Fiebrink(2013)]%
        {Fried2013}
\bibfield{author}{\bibinfo{person}{Ohad Fried} {and} \bibinfo{person}{Rebecca Fiebrink}.} \bibinfo{year}{2013}\natexlab{}.
\newblock \showarticletitle{Cross-modal Sound Mapping Using Deep Learning}. In \bibinfo{booktitle}{\emph{Proceedings of the International Conference on New Interfaces for Musical Expression}}. \bibinfo{publisher}{Graduate School of Culture Technology, KAIST}, \bibinfo{address}{Daejeon, Republic of Korea}, \bibinfo{pages}{531--534}.
\newblock
\showISSN{2220-4806}
\urldef\tempurl%
\url{https://doi.org/10.5281/zenodo.1178528}
\showDOI{\tempurl}


\bibitem[Gurevich and von Muehlen(2020)]%
        {gurevich2020accordiatron}
\bibfield{author}{\bibinfo{person}{Michael Gurevich} {and} \bibinfo{person}{Stephan von Muehlen}.} \bibinfo{year}{2020}\natexlab{}.
\newblock \showarticletitle{The Accordiatron: A MIDI controller for interactive music}.
\newblock \bibinfo{journal}{\emph{arXiv preprint arXiv:2010.01574}} (\bibinfo{year}{2020}).
\newblock


\bibitem[Ho and Salimans(2021)]%
        {ho_classifier-free_2021}
\bibfield{author}{\bibinfo{person}{Jonathan Ho} {and} \bibinfo{person}{Tim Salimans}.} \bibinfo{year}{2021}\natexlab{}.
\newblock \showarticletitle{Classifier-{Free} {Diffusion} {Guidance}}. In \bibinfo{booktitle}{\emph{{NeurIPS} 2021 {Workshop} on {Deep} {Generative} {Models} and {Downstream} {Applications}}}.
\newblock
\urldef\tempurl%
\url{https://openreview.net/forum?id=qw8AKxfYbI}
\showURL{%
\tempurl}


\bibitem[Huang et~al\mbox{.}(2014)]%
        {10.1145/2557500.2557544}
\bibfield{author}{\bibinfo{person}{Cheng-Zhi~Anna Huang}, \bibinfo{person}{David Duvenaud}, \bibinfo{person}{Kenneth~C. Arnold}, \bibinfo{person}{Brenton Partridge}, \bibinfo{person}{Josiah~W. Oberholtzer}, {and} \bibinfo{person}{Krzysztof~Z. Gajos}.} \bibinfo{year}{2014}\natexlab{}.
\newblock \showarticletitle{Active learning of intuitive control knobs for synthesizers using gaussian processes}. In \bibinfo{booktitle}{\emph{Proceedings of the 19th International Conference on Intelligent User Interfaces}} (Haifa, Israel) \emph{(\bibinfo{series}{IUI '14})}. \bibinfo{publisher}{Association for Computing Machinery}, \bibinfo{address}{New York, NY, USA}, \bibinfo{pages}{115–124}.
\newblock
\showISBNx{9781450321846}
\urldef\tempurl%
\url{https://doi.org/10.1145/2557500.2557544}
\showDOI{\tempurl}


\bibitem[Liu et~al\mbox{.}(2024)]%
        {liu2024mindcube}
\bibfield{author}{\bibinfo{person}{Fangzheng Liu}, \bibinfo{person}{Don~Derek Haddad}, {and} \bibinfo{person}{Joe Paradiso}.} \bibinfo{year}{2024}\natexlab{}.
\newblock \showarticletitle{MindCube: an Interactive Device for Gauging Emotions}. In \bibinfo{booktitle}{\emph{Adjunct Proceedings of the 37th Annual ACM Symposium on User Interface Software and Technology}}. \bibinfo{pages}{1--2}.
\newblock


\bibitem[Rombach et~al\mbox{.}(2022)]%
        {rombach_high-resolution_2022}
\bibfield{author}{\bibinfo{person}{Robin Rombach}, \bibinfo{person}{Andreas Blattmann}, \bibinfo{person}{Dominik Lorenz}, \bibinfo{person}{Patrick Esser}, {and} \bibinfo{person}{Björn Ommer}.} \bibinfo{year}{2022}\natexlab{}.
\newblock \showarticletitle{High-{Resolution} {Image} {Synthesis} {With} {Latent} {Diffusion} {Models}}. In \bibinfo{booktitle}{\emph{Proceedings of the {IEEE}/{CVF} {Conference} on {Computer} {Vision} and {Pattern} {Recognition} ({CVPR})}}. \bibinfo{pages}{10684--10695}.
\newblock


\bibitem[Scurto and Postel(2023)]%
        {scurto_soundwalking_2023}
\bibfield{author}{\bibinfo{person}{Hugo Scurto} {and} \bibinfo{person}{Ludmila Postel}.} \bibinfo{year}{2023}\natexlab{}.
\newblock \showarticletitle{Soundwalking {Deep} {Latent} {Spaces}}. In \bibinfo{booktitle}{\emph{Proceedings of the {International} {Conference} on {New} {Interfaces} for {Musical} {Expression}}}, \bibfield{editor}{\bibinfo{person}{Miguel Ortiz} {and} \bibinfo{person}{Adnan Marquez-Borbon}} (Eds.). \bibinfo{address}{Mexico City, Mexico}, \bibinfo{pages}{232--235}.
\newblock
\urldef\tempurl%
\url{https://doi.org/10.5281/zenodo.11189166}
\showDOI{\tempurl}
\newblock
\shownote{ISSN: 2220-4806}.


\bibitem[Shepardson and Magnusson(2023)]%
        {nime2023_32}
\bibfield{author}{\bibinfo{person}{Victor Shepardson} {and} \bibinfo{person}{Thor Magnusson}.} \bibinfo{year}{2023}\natexlab{}.
\newblock \showarticletitle{The Living Looper: Rethinking the Musical Loop as a Machine Action-Perception Loop}. In \bibinfo{booktitle}{\emph{Proceedings of the International Conference on New Interfaces for Musical Expression}}, \bibfield{editor}{\bibinfo{person}{Miguel Ortiz} {and} \bibinfo{person}{Adnan Marquez-Borbon}} (Eds.). \bibinfo{address}{Mexico City, Mexico}, Article \bibinfo{articleno}{32}, \bibinfo{numpages}{8}~pages.
\newblock
\showISSN{2220-4806}
\urldef\tempurl%
\url{https://doi.org/10.5281/zenodo.11189164}
\showDOI{\tempurl}


\bibitem[Torre et~al\mbox{.}(2016)]%
        {torre2016hands}
\bibfield{author}{\bibinfo{person}{Giuseppe Torre}, \bibinfo{person}{Kristina Andersen}, {and} \bibinfo{person}{Frank Bald{\'e}}.} \bibinfo{year}{2016}\natexlab{}.
\newblock \showarticletitle{The Hands: The making of a digital musical instrument}.
\newblock \bibinfo{journal}{\emph{Computer Music Journal}} \bibinfo{volume}{40}, \bibinfo{number}{2} (\bibinfo{year}{2016}), \bibinfo{pages}{22--34}.
\newblock


\bibitem[Trolland et~al\mbox{.}(2022)]%
        {trolland2022airsticks}
\bibfield{author}{\bibinfo{person}{Sam Trolland}, \bibinfo{person}{Alon Ilsar}, \bibinfo{person}{Ciaran Frame}, \bibinfo{person}{Jon McCormack}, {and} \bibinfo{person}{Elliott Wilson}.} \bibinfo{year}{2022}\natexlab{}.
\newblock \showarticletitle{AirSticks 2.0: Instrument design for expressive gestural interaction}. In \bibinfo{booktitle}{\emph{NIME 2022}}. PubPub.
\newblock


\bibitem[Wong et~al\mbox{.}(2008)]%
        {wong2008designing}
\bibfield{author}{\bibinfo{person}{Elaine~L Wong}, \bibinfo{person}{Wilson~YF Yuen}, {and} \bibinfo{person}{Clifford~ST Choy}.} \bibinfo{year}{2008}\natexlab{}.
\newblock \showarticletitle{Designing wii controller: a powerful musical instrument in an interactive music performance system}. In \bibinfo{booktitle}{\emph{Proceedings of the 6th International Conference on Advances in Mobile Computing and Multimedia}}. \bibinfo{pages}{82--87}.
\newblock


\bibitem[Woodward and Kanjo(2020)]%
        {woodward2020ifidgetcube}
\bibfield{author}{\bibinfo{person}{Kieran Woodward} {and} \bibinfo{person}{Eiman Kanjo}.} \bibinfo{year}{2020}\natexlab{}.
\newblock \showarticletitle{ifidgetcube: Tangible fidgeting interfaces (tfis) to monitor and improve mental wellbeing}.
\newblock \bibinfo{journal}{\emph{IEEE Sensors Journal}} \bibinfo{volume}{21}, \bibinfo{number}{13} (\bibinfo{year}{2020}), \bibinfo{pages}{14300--14307}.
\newblock


\end{thebibliography}

\end{document}